\documentstyle[11pt]{article}
\def\lb{\label}
\def\be{\begin{equation}}
\def\ee{\end{equation}}
\def\ba{\begin{eqnarray}}
\def\ea{\end{eqnarray}}
\def\ds{\displaystyle}
\def\ol{\overline}

\def\p{\partial_}

\renewcommand{\theequation}{\arabic{section}.\arabic{equation}}
\begin{document}

\title{
   \begin{flushright} \begin{small}
     LAPTH-833/01  \\ DTP-MSU/01-03 \\  
  \end{small} \end{flushright}
{\bf Bertotti-Robinson type solutions to Dilaton-Axion Gravity} } 
\author{
	G\'erard Cl\'ement$^{a}$
\thanks{Email: gclement@lapp.in2p3.fr}
 and Dmitri Gal'tsov$^{a,b}$
\thanks{Email: galtsov@lapp.in2p3.fr and galtsov@grg.phys.msu.su} \\
$^{a}$Laboratoire de  Physique Th\'eorique LAPTH (CNRS), \\
B.P.110, F-74941 Annecy-le-Vieux cedex, France\\
$^{b}$Department of Theoretical Physics,
	 Moscow State University,\\ 119899, Moscow, Russia,}

\date{6 February 2001}
\maketitle
\begin{abstract}
We present a new solution to dilaton-axion gravity which looks like a
rotating  Bertotti-Robinson (BR) Universe. It is supported by an
homogeneous Maxwell field and a linear  axion and can be obtained as a
near-horizon limit of extremal rotating dilaton-axion black holes. It
has the isometry $SL(2,R)\times U(1)$ where $U(1)$ is the remnant of
the $SO(3)$ symmetry of BR broken by rotation, while $SL(2,R)$
corresponds to the $AdS_2$ sector which no longer factors out of the
full spacetime. Alternatively our solution can be obtained from the
$D=5$ vacuum counterpart to the dyonic BR with equal electric and
magnetic field strengths. The derivation amounts to smearing it in
$D=6$ and then reducing to $D=4$ with dualization of one Kaluza-Klein
two-form in $D=5$ to produce an axion.  Using a similar dualization
procedure, the rotating BR solution is uplifted to  $D=11$
supergravity. We show that it breaks all supersymmetries of $N=4$
supergravity in $D=4$, and that its higher dimensional embeddings are
not supersymmetric either. But, hopefully  it may provide a new arena
for corformal mechanics and holography. Applying a complex coordinate
transformation we also derive a BR solution endowed with a NUT
parameter.

\end{abstract}   

\section{Introduction}  The discovery of  AdS/CFT
dualities \cite{Ma98} (for a review see \cite{Ma5}) stimulated search for
geometries containing AdS sectors. Recall that AdS
geometry typically arises as the near horizon (throat) limit of static
charged BPS black holes and/or p-branes in various dimensions. The
known examples of  ADS/CFT correspondence make use of geometries
$AdS_n\times K$ with $K$ being some compact manifold. For rotating black
holes/branes the near horizon limit generically is different: one finds
a non-trivial mixing of $AdS_n$ and $K$. Nevertheless, the asymptotic
geometry relevant for holography may remain unaffected by rotation for
$n\geq 3$, so the ADS/CFT correspondence applies
directly. This is not so in the case of $AdS_2$
\cite{BaHo99,La99}.  While for non-rotating extremal charged black
holes the near-horizon geometry factorizes as $AdS_2\times S^2$, in the
rotating case the azimuthal coordinates mixes with the time coordinate in such
a way that the $AdS_2\times S^2$ geometry is not recovered asymptotically.
Therefore, in higher dimensions the rotation parameter just adds a specific
excitation mode in the dual theory  \cite{CvLa98,HaHuTa99,CvLa99,BePa99},
but in four dimensions it apparently leads to more seriours consequencies,
whose nature is not clear yet. Additional problems in the
four-dimensional case are related to the fact that the  $AdS_2$ 
holography is less well understood than the higher-dimensional examples
\cite{St98,St99,CaMi99,GiTo99,NaNa99} (for some recent progress in this
direction see \cite{CaCaKlMi00}).

Bardeen and Horowitz \cite{BaHo99} observed that violation of the direct
product structure $AdS_2\times K$ due to rotation in $D=4$ is manifest
already for vacuum Kerr black holes. In the throat geometry  of the extreme
Kerr  (and Kerr-Newman)four-dimensional black hole  
one finds the mixing of azimuthal and time coordinates which does not
disappear in the asymptotic region, but  grows infinitely instead leading to
the singular nature  of the conformal boundary. Nevertheless, the geometry
still share some important features with  $AdS_n\times K$ spacetimes such
as (partial) confinement of timelike geodesics and discreetness of the
Klein-Gordon particle spectrum on the  geodesically complete AdS patch. But,
apart from the fact that the near-horizon spacetime is no longer the direct
product of  $AdS_2$ with something, the geometry is also plagued by cumbersome
$\theta$-depending factors which modify the spectrum of the angular
Laplacian. It turns out that the spectrum of the Klein-Gordon field
contains a continuous sector which exhibits superradiance
inherited from Kerr. All this substantially complicates the analysis
and no definitive conclusion was gained in  \cite{BaHo99} about the possible
relevance of such geometries in holography.

Here we present another geometry containing the $AdS_2$ sector mixed
non-trivially with the rest of the spacetime which  has the advantage of not
being afflicted by $\theta$-factors. It
can be obtained as the  near-horizon limit of extremal rotating
dilaton-axion black holes (solutions to the Einstein-Maxwell-dilaton-axion
(EMDA) theory). This is therefore a non-vacuum solution which is supported by
a homogeneous Maxwell field (similarly to the Bertotti-Robinson (BR) spacetime)
and a linear axion. The rotation breaks the $SO(3)$ symmetry of BR so that the
solution is only axially symmetric.
Meanwhile, as in the Bardeen-Horowitz case, the $SL(2,R)$ symmetry of the
AdS component still holds, so the full isometry group is $SL(2,R)\times U(1)$.

We show that this geometry has a non-trivial connection with BR via a
non-local  dimensional reduction mechanism \cite{ChGaMaSh99} 
involving dualization of Kaluza-Klein two-forms in order to generate
higher-rank antisymmetric forms. Starting with the dyonic $D=4$ BR with equal
strengths of the electric and magnetic components one finds its purely vacuum
$D=5$ counterpart (the KK dilaton is not excited in the symmetric
dyon case). This solution can be smeared into the sixth dimension providing the
`BR6' vacuum geometry. Then one performs dimensional reduction back to five
dimensions dualizing the Kaluza-Klein two-form and reinterpreting the resulting
three-form as a NS-NS field. Finally going to four dimensions via the usual KK
reduction one recovers the EMDA theory counterpart of the BR6 which
coincides with our near-horizon limit of  rotating EMDA black holes.

Using the same mechanism of generation of antisymmetric forms, we uplift the
new solution into eleven-dimensional supergravity where it is supported by
a non-trivial four-form. This is based on the correspondence
between eight-dimensional vacuum gravity with two commuting
Killing vector fields and a consistent $2+3+6$ three-block truncation of 
$D=11$ supergravity \cite{ChGaSh00}. To apply this technique one has first to
smear BR6  in two additional dimensions and then use dualization of the
KK two-form in six dimensions to get the four-form which is then used to
reconstruct the four-form of $D=11$ supergravity. In doing this there are
two options in the choice of the Killing vectors which lead to two 
supergravity solutions with different four-forms but the same metric.

Since rotation breaks the BPS condition for extremal dilaton-axion black holes,
it can be expected that our rotating BR solution is not supersymmetric in the
sense of $D=4,\,{\cal N}=4$ supergravity. To check this, we consider
the purely algebraic equation for variation of dilatino and show that this
variation is non-zero. We also check by a direct computation that the $D=11$
embedding of our solution is not supersymmetric in the sense of $D=11$
supergravity. Nevertheless we argue that the rotating BR geometry provides
an interesting new arena for conformal mechanics and holography. 

\setcounter{equation}{0}
\section{Near-horizon limit of  rotating EMDA black holes}
Consider the Einstein-Maxwell-Dilaton-Axion (EMDA) theory which is a
truncated version of   the bosonic sector of $D=4,\;{\cal N}=4$ supergravity.
The action describes  
the gravity--coupled system of two scalar fields: dilaton $\phi$ and 
(pseudoscalar) axion $\kappa$, and an Abelian vector field $A_{\mu}$ : 
\be  \lb{ac} S = \frac{1}{16\pi}\int d^4x\sqrt{|g|}\left\{-R + 
2\partial_\mu\phi\partial^\mu\phi + \frac{1}{2} e^{4\phi} 
{\partial_\mu}\kappa\partial^\mu\kappa -e^{-2\phi}F_{\mu\nu}F^{\mu\nu}-\kappa 
F_{\mu\nu}{\tilde F}^{\mu\nu}\right\} , 
\ee 
where\footnote{Here $E^{\mu\nu\lambda\tau} \equiv 
|g|^{-1/2}\varepsilon^{\mu\nu\lambda\tau}$, with 
$\varepsilon^{1234} = +1$, where $x^4 = t$ is the time 
coordinate.} ${\tilde 
F}^{\mu\nu}=\frac{1}{2}E^{\mu\nu\lambda\tau}F_{\lambda\tau},\; 
F=dA\;$. The black hole solutions to this theory were extensively studied in 
the recent past \cite{Ka92,Ka94,GaKe94,GaKe96,BeKaOr96,ClGa96}. 
Thesedepend on six real parameters, the complex mass ${\cal M} = M +
iN$ (where $N$ is the NUT parameter), electromagnetic charge
${\cal Q} = Q + iP$ and axion-dilaton charge ${\cal D} = D + iA$
constrained by ${\cal D} = - {\cal Q}^2/2{\cal M}$, and the
rotation parameter $a$. The general black hole metric is of the
form
\be\lb{GK}
ds^2 = \frac{\Delta - a^2\sin^2\theta}{\Sigma}\, (dt - \omega\,d\varphi)^2 -
\Sigma \left( \frac{dr^2}{\Delta} + d\theta^2 +
\frac{\Delta\sin^2\theta}{\Delta - a^2\sin^2\theta}\,d\varphi^2 \right) \,,
\ee
with
\ba
\Delta & = & (r - r_-)(r - 2M) +a^2 - (N -N_-)^2\,, \nonumber \\
\Sigma & = & r(r - r_-) + (a\cos\theta + N)^2 - N_-^2\,,
\\ \omega & = & \frac{2}{a^2\sin^2\theta - \Delta} [N\Delta\cos\theta +
a\sin^2\theta\,(M(r - r_-) + N(N - N_-))] \,, \nonumber
\ea
and
\be
r_- = \frac{M|{\cal Q}|^2}{|{\cal M}|^2} \,, \quad N_- = \frac{N}{2M}\,r_-\,.
\ee
The vector field may be parametrized by two scalar electric ($v$) and
magnetic ($u$) potentials defined by
\ba\lb{Fpots}
F_{i0} & = & \partial_i v/\sqrt{2} \,, \nonumber \\
e^{-2\phi}F^{ij} + \kappa \tilde{F}^{ij} & = &
(\Sigma\sin\theta)^{-1} \epsilon^{ijk}
\partial_k u/\sqrt{2} \,.
\ea
The potentials $v$ and $u$ and the axion-dilaton field are given
(after adapting the formulas of \cite{GaKe94} to the conventions of
\cite{GaKe96}) by
\ba\lb{pots}
v & = & \sqrt{2}\,\frac{e^{\phi_{\infty}}}{\Sigma}\,{\rm Re}[{\cal
Q}(r-r_- - i\delta)] \,, \nonumber \\ u & = & \sqrt{2}\,
\frac{e^{\phi_{\infty}}}{\Sigma}\,{\rm Re}[{\cal Q}z_{\infty}
(r-r_- - i\delta)] \,, \nonumber \\ z & \equiv & \kappa +
i\,e^{-2\phi} = \frac{z_{\infty}\rho + {\cal
D}^*z_{\infty}^*}{\rho + {\cal D}^*}\,,
\ea
where
\be
\delta = a\cos\theta + N - N_- \,, \quad \rho = r - \frac{{\cal
M}r_-} {2M} - i\delta\,,
\ee
and the (physically irrelevant) asymptotic value of the
axion-dilaton field $z_{\infty} \equiv \kappa_{\infty} +
i\,e^{-2\phi_{\infty}}$ will be chosen later for convenience.

The metric (\ref{GK}) has two horizons located at the zeroes $r =
r_H^{\pm}$ of the function $\Delta$:
\be
r_H^{\pm} = M + r_-/2 \pm \sqrt{(|{\cal M}| - |{\cal D}|)^2 - a^2}\,.
\ee
The extremal solutions correspond to the case
\be
|{\cal D}| = |{\cal M}| - a
\ee
where these two horizons coincide, with $\Sigma > 0$ (without loss
of generality we assume $a > 0$). In this case, using
\begin{equation}
r - r_{-} = r - r_H + a\frac{M}{|{\cal M}|}\,,
\end{equation}
we can rewrite the metric as
\be\lb{extrds}
ds^2 = \frac{\Sigma\Delta\sin^2\theta}{\Gamma}\,dt^2 -
\Sigma(\frac{dr^2}{\Delta} + d\theta^2) -
\frac{\Gamma}{\Sigma}(d\varphi - \Omega dt)^2 \,
\ee
with
\ba\lb{extr}
\Gamma & = & \Delta\,[(\eta + 4\gamma)\sin^2\theta -
4N^2\cos^2\theta] + 4(\gamma- aN\cos\theta)^2\sin^2\theta \,,
\nonumber \\ \Sigma & = & \eta + 2\gamma\,, \quad \Omega  =
2\,(N\eta\cos\theta + a\gamma\sin^2\theta)/\Gamma\,, \\ \Delta & =
& (r - r_H)^ 2\,, \quad \eta = \Delta - a^2\sin^2\theta\,, \quad
\gamma = M(r - r_H) + a(|{\cal M}| + N\cos\theta)\,. \nonumber
\ea

On the horizon $r = r_H$, the metric functions occurring in (\ref{extr})
simplify to
\ba
\Gamma_H & = & 4a^2|{\cal M}|^2\sin^2\theta\,,  \quad \Omega_H =
1/2|{\cal M}|\,, \nonumber
\\ \quad \Sigma_H & = & 2a(|{\cal M}| + N\cos\theta) -
a^2\sin^2\theta\,.
\ea
It follows that in the static case $a = 0$, the horizon reduces to a point,
so that the question of near-horizon limit becomes meaningless.But for
rotating extremal black holes $a \neq 0$ one finds a non-triviallimiting
solution. In this case, before taking the near-horizon limit, let us first
transform to a frame co-rotating with the horizon:
\be
\overline{\varphi} \equiv \varphi - \Omega_H\,t\,.
\ee
In this frame, the differential angular velocity is, near the horizon,
\be
\overline{\Omega} = \Omega - \Omega_H =
-\frac{2aM(r-r_H)\sin^2\theta}{\Gamma} + O(\Delta)\,.
\ee
Let us now put
\be\lb{newcoord}
r - r_H \equiv \lambda x\,, \quad \cos\theta\equiv y\,, \quad t
\equiv \frac{r_0^2}{\lambda}\overline{t} \qquad (r_0^2 = 2a|{\cal
M}|)\,,
\ee
and take the limit $\lambda \to 0$. In this limit the extreme metric in the
rotating frame reduces to
\ba\lb{nhna}
ds^2 & = & r_0^2 \,[ (\alpha + \nu y + \beta y^2)
(x^2\,dt^2 - \frac{dx^2}{x^2}) \nonumber
\\ & & - \frac{\alpha + \nu y + \beta y^2}{1 - y^2}\,dy^2 - \frac{1 -
y^2}{\alpha + \nu y + \beta y^2}(d\varphi + \mu x\,dt)^2 ]\,,
\ea
where $\beta = a/2|{\cal M}|$, $\alpha = 1 -
\beta$, $\mu = M/|{\cal M}|$, $\nu = N/|{\cal M}|$ ($\mu^2 + \nu^2
= 1$), and we have relabelled the coordinates
$(\overline{t},\overline{\varphi}) \to (t, \varphi)$.

Just as the extreme Kerr or Kerr-Newman geometries in
Einstein-Maxwell theory \cite{BaHo99}, the near-horizon geometry
(\ref{nh}) admits four Killing vectors
\ba
L_1 & = & x\partial_x - t\partial_t \,, \nonumber \\ L_2 & = &
xt\partial_x - \frac{1}{2}(x^{-2} + t^{2})\partial_t + \mu
x^{-1}\partial_{\varphi}\,, \nonumber \\ L_3 & = & \partial_t\,,
\\ L_4 & = & \partial_{\varphi}\,, \nonumber
\ea
generating the group $SL(2,R) \times U(1)$. Indeed, the metric
(\ref{nhna}) becomes, in the NUT-less case $\nu = 0$,
\be\lb{nha}
ds^2 = r_0^2 \, [ \,(\alpha + \beta y^2) (x^2\,dt^2 -
\frac{dx^2}{x^2}) - \frac{\alpha + \beta y^2}{1 - y^2}\,dy^2 -
\frac{1 - y^2}{\alpha + \beta y^2}(d\varphi + x\,dt)^2\,]\,.
\ee
This is similar in form to the extreme Kerr-Newman
near-horizon metric \cite{BaHo99,Za98}
\ba\lb{nhkn}
ds_{EM}^2 & = & r_{0EM}^2\,[\,(\alpha_{EM} + \beta_{EM} y^2)
(x^2\,dt^2 - \frac{dx^2}{x^2})  - \frac{\alpha_{EM} + \beta_{EM}
y^2}{1 - y^2}\,dy^2 \nonumber
\\ & & - \frac{1 - y^2}{\alpha_{EM} + \beta_{EM}
y^2}(d\varphi + \mu_{EM}x\,dt)^2\,]
\ea
with  $r_{0EM}^2 = M^2 + a^2$, $\beta_{EM} = a^2/(M^2 + a^2)$,
$\alpha_{EM} = 1 -
\beta_{EM}$, $\mu_{EM} = 2aM/(M^2 + a^2)$, where $M^2 = a^2 +
¦{\cal Q}¦^2$. The two extreme geometries (\ref{nha}) and
(\ref{nhkn}) become identical in the neutral case $¦{\cal Q}¦ = 0$
($M = a$) and reduce to the extreme Kerr geometry with $\alpha =
\beta = 1/2$.

Now let us take the limit $a \to 0$ in the NUT-less
near-horizon geometry  (\ref{nha}), while keeping $r_0^2 = 2aM$
fixed. In this manner we arrive at the metric
\be\lb{nh}
ds^2 = x^2\,dt^2 - \frac{dx^2}{x^2} - \frac{dy^2}{1 - y^2} - (1 -
y^2)(d\varphi + x\,dt)^2
\ee
(we have scaled $r_0^2$ to unity).
This metric is remarkably similar in form to the Bertotti-Robinson metric
(the near-horizon geometry of the extreme Reissner-Nordstr\"{o}m black hole,
i.e. the static limit $a \to 0$ of (\ref{nhkn})),
\be\lb{br}
ds^2 = x^2\,dt^2 - \frac{dx^2}{x^2} - \frac{dy^2}{1 - y^2} 
- (1 -y^2)\,d\varphi^2.
\ee
However, unlike the BR metric, the BREMDA metric (\ref{nh}) is
non-static, and invariant only under the group $SL(2,R) \times
U(1)$.  
The $t, x$ coordinates do not cover the full AdS hyperboloid. The
geodesically complete manifold is covered by another coordinate patch in
which case the metric reads (we preserve the same symbols for radial and
azimuthal coordinates)
\be\lb{nhz}
ds^2 = (1+x^2)\,d\tau^2 - \frac{dx^2}{1+x^2} - \frac{dy^2}{1 - y^2}-
(1 -y^2)\,(d\varphi +xd\tau)^2.
\ee
Another useful coordinate system (also incomplete) is given by 
\be\lb{nh-}
ds^2 = (x^2-1)\,d\tau^2 - \frac{dx^2}{x^2-1} - d\theta^2 -\frac{dy^2}{1 - y^2}-
(1 -y^2)\,(d\varphi +xd\tau)^2.
\ee

Now proceed along the same lines with the matter fields. From the last
Eq. (\ref{pots}), the dilaton and axion fields reduce on the horizon to
\ba
e^{-2\phi_H} & = & e^{-2\phi_{\infty}}\,\frac{2a(|{\cal M}| + Ny)
- a^2(1 - y^2)} {(M + D)^2 + (N + A + ay)^2}\,, \nonumber \\
\kappa_H & = & \kappa_{\infty} + 2e^{-2\phi_{\infty}}\,\frac{D(N +
ay) - AM} {(M + D)^2 + (N + A + ay)^2}\,.
\ea
For $N = 0$ and in the limit $a \to 0$, these become (after neglecting terms
of order $a^2$)
\ba
e^{-2\phi_H} & = & e^{-2\phi_{\infty}}\,\frac{aM}{P^2}\,,
\nonumber
\\ \kappa_H & = & \kappa_{\infty} -
e^{-2\phi_{\infty}}\,\frac{M}{P^2}(A - ay)\,.
\ea
Now we choose for convenience
\be\lb{zinf}
z_{\infty} = -2P{\cal Q}^*/r_0^2
\ee
($r_0^2 = 2aM$), leading to the BREMDA dilaton and axion fields
\be\lb{ad}
\phi_H = 0,\quad \kappa_H = \cos\theta,
\ee
 irrespective of the original
values of $Q$ and $P$ (provided $P \neq 0$). 

The determination of the near-horizon behavior of the gauge field
is more involved. With the choice (\ref{zinf}), the scalar
potentials for the NUT-less extreme black hole are
\ba
v & = & \sqrt{2}\,\frac{e^{\phi_{\infty}}}{\Sigma}\,(Q(r-r_{H}+a)
+ Pa\cos\theta) \,, \nonumber
\\ u & = & -\sqrt{2}\,
\frac{e^{-\phi_{\infty}}}{\Sigma}\,\frac{2M(M-a)}{P}\,(r-r_{H}+a)\,.
\ea
First, we must transform these to the rotating and rescaled
coordinate frame ($\overline{\varphi},\overline{t}$)
\be
\left(\begin{array}{c}
  d\overline{\varphi} \\
  d\overline{t}
\end{array}\right)
=
\left(\begin{array}{cc}
  1 & -\Omega_{H} \\
  0 & \lambda/r_0^2
\end{array}\right)
\left(\begin{array}{c}
  d\varphi \\
  dt
\end{array}\right)\,.
\ee
 From (\ref{Fpots}) we obtain the transformation laws
\ba\lb{transfpots}
\partial_i \overline{v} & = & \frac{\ds r_0^2/\lambda}{\ds \Sigma^2\Delta\sin^2\theta/\Gamma^2-\Omega^2}
\Big\{
\left(\frac{(\Sigma^2\Delta\sin^2\theta}{\Gamma^2}-\Omega\overline{\Omega}\right)
\partial_i v \nonumber \\
& & \quad + \frac{\Omega_H
e^{2\phi}\Delta\sin\theta}{\Gamma}\,g_{ij}\epsilon^{jk}(\partial_k
u - \kappa\partial_k v) \Big\}\,, \nonumber \\
\partial_i\overline{u} & = & \frac{\ds r_0^2/\lambda}{\ds \Sigma^2\Delta\sin^2\theta/\Gamma^2-\Omega^2}
\Big\{\left(\frac{(\Sigma^2\Delta\sin^2\theta}{\Gamma^2}-\Omega\overline{\Omega}\right)
\partial_i u \\
& & \quad - \frac{\Omega_H
e^{2\phi}\Delta\sin\theta}{\Gamma}\,g_{ij}\epsilon^{jk}((e^{-4\phi}+\kappa^2)\partial_k
v - \kappa\partial_k u) \Big\} \nonumber
\ea
($i,j = r,\theta$). Then, using
\be
\partial_r\Sigma  \simeq 2M\,, \quad \partial_{\theta}\Sigma  =
-2a^2\sin\theta\cos\theta\,, \ee we evaluate the derivatives in
(\ref{transfpots}) near the horizon, keeping only the leading
terms in $a$ (only the partial derivatives relative to $\theta$
contribute in this order):
\ba
\partial_r \overline{v}_H & \simeq -
{\ds\frac{r_0}{\lambda}}\cos\theta \,, \quad & \partial_\theta
\overline{v}_H \simeq \frac{r_0}{\lambda} (r-r_H)\sin\theta \,,
\nonumber \\ \partial_r \overline{u}_H & \simeq
{\ds\frac{r_0}{\lambda}}\sin^2\theta \,, \quad &
\partial_\theta \overline{u}_H \simeq \frac{r_0}{\lambda}
(r-r_H)2\sin\theta\cos\theta \,.
\ea
 From these we obtain the near-horizon potentials
\ba\lb{nhuv}
\overline{v}_H & = & - r_0x\cos\theta \,, \nonumber \\ \overline{u}_H & = &
r_0x\sin^2\theta\,,
\ea
again irrespective of the original values of the electric and
magnetic charges. We have checked that these, together with the
near-horizon metric (\ref{nh}) and axion-dilaton fields (\ref{ad}) solve
the field equations as given in \cite{GaKe96} . From the potentials
(\ref{nhuv}), after setting $r_0$ to unity, we recover the near-horizon gauge
fields  
\ba\lb{F}  F_{14} = -\frac{y}{\sqrt{2}}\,, & \quad F_{23} =
-{\ds\frac{1}{\sqrt{2}}}\,, \quad & F_{24} = -\frac{x}{\sqrt{2}}\,, \nonumber
\\ F^{13} = -\frac{xy}{\sqrt{2}}\,, & \quad F^{14} =
{\ds\frac{y}{\sqrt{2}}}\,, \quad & F^{23} = -\frac{1}{\sqrt{2}} \ea (with
$x^1=x$ and $x^2=y = \cos\theta$), deriving from the gauge potentials $A_3 = 
-y/\sqrt{2}$, $A_4 = -xy/\sqrt{2}$ .Finally, passing to more
general coordinates containing a free parameter $b$, we can write the
BREMDA solution as  follows~\footnote{Let us here mention that in
\cite{JoMy95} the low-energy limit of a certain conformal field theory was
shown to correspond to a formal near-horizon limit of Kerr-NUT solutions of
EMDA, with metric and matter fields different from (\ref{bremda}). However one
can show that these fields (given in Eq. (3.4) of \cite{JoMy95}) do not solve
the field equations of dilaton-axion gravity.}  
\ba\label{bremda}  
ds^2&=&(x^2+b) d\tau^2-\frac{dx^2}{x^2+b}
-d\theta^2-\sin^2\theta (d\varphi+x d\tau)^2,\nonumber\\ A&=&A_\mu
dx^\mu=-\frac{\cos\theta}{\sqrt{2}}(d\varphi+x d\tau),\quad \kappa=\cos\theta.
\ea 
For $b=0$ this coincides with the solution derived above by the limiting
procedure (Poincar\'e coordinates on AdS sector), for positive non-zero $b$
(usually set to $b=1$) one has a coordinate patch covering the full AdS
hyperboloid. For comparison consider the near-horizon limit of the Kerr
solution \cite{BaHo99}. Setting in (\ref{nha}) $r_0=1, \alpha=\beta=1/2$ and
passing to similar coordinates, we obtain \be\lb{BHz}  
ds^2 = \frac12(1+\cos^2\theta)\left[(x^2+b)\,d\tau^2 - 
\frac{dx^2}{(x^2+b)} -  d\theta^2\right] -
\frac{2\sin^2\theta}{1+\cos^2\theta}  (d\phi+xd\tau)^2.  
\ee 
In both cases the mixing of the azimuthal and time coordinates does not
vanish as $x\to\infty$. Both metrics coincide in the equatorial plane, but
differ for $\theta \neq \pi/2$. The BREMDA geometry is simpler due to
the absence of cumbersome angular factors, and apparently is more suitable for
the search of a holographic dual. We return to this question in a separate
publication \cite{ClGa01}, while here we discuss other geometrical aspects of
the new solution related to its embeddings in higher dimensions.


\setcounter{equation}{0}
\section{$D=4$ EMDA from $D=6$ Einstein gravity}

Let us now show how $D=4$ EMDA theory can be derived from the purely vacuum
Einstein theory in six dimensions. This may be hinted from the following
considerations. Dimensional reduction of stationary $D=4$ EMDA to three
dimensions leads to a gravity coupled sigma-model with the target space
isometry group $Sp(4,R)$ \cite{Ga95,GaKe95}, while dimensional reduction of
the 6D vacuum gravity to three dimensions gives a sigma model with the
$SL(4,R)$ target space symmetry \cite{Ma79}. It was shown in
\cite{ChGaMaSh99} that   a consistent truncation of the $SL(4,R)$ sigma model
to the $Sp(4,R)$ one exists, i.e. any
stationary solution of $D=4$  EMDA gravity has a $D=6$ vacuum gravity
counterpart and vice versa. Here we show that this holds not only for
stationary, but for any solutions of two theories. This duality is
essentially non-local: it involves dualization of the Kaluza-Klein
two-form in the intermediate five dimensions.

Let us start with  the action 
\be S = - \int d^6x \sqrt{|g_6|} R_6 \,, 
\ee 
denoting the 6-dimensional coordinates as $x^\mu, \;\eta,\;\chi$, and make the
assumption of two commuting spacelike Killing vectors $\partial_{\eta}$,
$\partial_{\chi}$. In any number of dimensions, the Kaluza-Klein dimensional
reduction
\be\lb{redn}
ds_{n+1}^2 = e^{-2c\hat{\phi}}ds_n^2 - e^{2(n-2)c\hat{\phi}}(d\eta
+ \hat{C}_{\mu}dx^{\mu})^2
\ee
gives
\be
\sqrt{|g_{n+1}|}R_{n+1} = \sqrt{|g_{n}|}\,[R_n -
(n-1)(n-2)c^2(\partial\hat{\phi})^2 +
\frac{1}{4}e^{2(n-1)c\hat{\phi}}F^2(\hat{C}) +
2c\nabla^2\hat{\phi}]\,.
\ee
For $n = 5$, $c = 1/\sqrt{6}$, this leads, after dualizing the
2-form $F(\hat{C})$ to a 3-form $\hat{H} = d\hat{K}$,
\be\lb{hatH}
F^{\mu\nu}(\hat{C}) = -\frac{1}{6\sqrt{|g_{5}|}}\,
e^{-\alpha\hat{\phi}}
\epsilon^{\mu\nu\lambda\rho\sigma}\,\hat{H}_{\lambda\rho\sigma}\,,
\ee
to the reduced action for 6-dimensional sourceless gravity
\be\lb{ac5}
S_5 = \int d^5x \sqrt{|g_5|}[-R_5 + 2(\partial\hat{\phi})^2 +
\frac{1}{12}e^{-\alpha\hat{\phi}}\hat{H}^2]\,,
\ee
with $\alpha = 4\sqrt{2/3}$.

In (\ref{ac5}) we recognize the action of 5-dimensional gravity
coupled to a dilaton and a 3-form, as written down in \cite{ChGaMaSh99}.
In this paper it was observed that, under the assumption of two
commuting Killing vectors $\partial_4$ and $\partial_5$, this
theory reduces to a 3-dimensional $\sigma$ model with the
$SL(4,R)$ symmetry group. Here we see that this symmetry follows
directly from the $SL(4,R)$ symmetry of sourceless $D=6$ gravity with
3 Killing vectors, which is a special case of sourceless
n-dimensional gravity with $(n-3)$ Killing vectors, as discussed by
Maison \cite{Ma79}.

In a second step, the 5-dimensional theory (\ref{ac5}) with a
spacelike Killing vector $\partial\chi$ is further reduced by the
Kaluza-Klein ansatz
\ba\lb{red4a}
ds_5^2 & = & e^{-2\sigma/\sqrt{3}}\,ds_4^2 -
e^{4\sigma/\sqrt{3}}\,(d\chi + D_{\mu}\,dx^{\mu})^2\,, \nonumber
\\ \hat{K} & = & K_{\mu\nu}\,dx^{\mu}\wedge dx^{\nu} +
E_{\mu}\,d\chi\wedge dx^{\mu}\,,
\ea
to the action
\ba\lb{ac4}
S_4 & = & \int d^4x \sqrt{|g_4|}[-R_4 + 2(\partial\phi)^2 +
(\partial\psi)^2 + \frac{1}{2}e^{4\phi}(\partial\kappa)^2 -
\frac{1}{4}e^{2(\psi-\phi)}F^2(D) \nonumber \\ & &
-\frac{1}{4}e^{-2(\psi+\phi)}F^2(E) -
\frac{\kappa}{4}(F(D)\tilde{F}(E) +
F(E)\tilde{F}(D))]\,,
\ea
where
\be
\phi = \sqrt{\frac{2}{3}}\,\hat{\phi} -
\sqrt{\frac{1}{3}}\,\sigma\,, \quad \psi =
\sqrt{2}\left(\sqrt{\frac{1}{3}}\,\hat{\phi} +
\sqrt{\frac{2}{3}}\,\sigma\right)\,,
\ee
and $\kappa$ is the dual of the 3-form
\be\lb{H}
H \equiv dK - D\wedge F(E) = - e^{4\phi}\,*d\kappa\,.
\ee

Remarkably, the equations of motion for the fields $D$, $E$ and $\psi$
deriving from the action (\ref{ac4})
\ba\lb{DEpsi}
\nabla_{\mu}(e^{2(\psi-\phi)}F^{\mu\nu}(D) +
\kappa\tilde{F}^{\mu\nu}(E)) & = & 0\,, \nonumber \\
\nabla_{\mu}(e^{-2(\psi+\phi)}F^{\mu\nu}(E) +
\kappa\tilde{F}^{\mu\nu}(D)) & = & 0\,, \\ \nabla^2\psi +
\frac{1}{4}e^{-2\phi}(e^{2\psi}F^2(D) - e^{-2\psi}F^2(E)) & = & 0
\nonumber
\ea
are consistent with the ansatz
\be\lb{an}
\psi = 0\,, \qquad D_{\mu} = E_{\mu} \equiv \sqrt{2}A_{\mu}\,,
\ee
which reduces the action (\ref{ac4}) to the action (\ref{ac}) of
EMDA (this is similar to the reduction of five-dimensional
Kaluza-Klein theory to self-dual Einstein-Maxwell theory).

This two-step reduction of 6-dimensional vacuum gravity can be
summarized in a direct reduction from 6 to 4 dimensions. From
(\ref{hatH}),
\be
F_{\mu5}(\hat{C}) = \partial_{\mu}\hat{C}_5 =
-e^{-4\phi}\tilde{H}_{\mu} = \partial_{\mu}\kappa\,,
\ee
where $\mu = 1, \ldots, 4$, so that the 5-dimensional 1-form
$\hat{C}$ reduces according to
\be
\hat{C}= C_{\mu}\,dx^{\mu} + \kappa\,d\chi\,.
\ee
The two successive Kaluza-Klein ans\"atze (\ref{redn}) (for $n =
5$, $c = 1/\sqrt{6}$) and (\ref{red4a}) can be combined into
\be\lb{red64}
ds_6^2 = e^{-\psi}\,ds_4^2 - e^{\psi-2\phi}\,(d\chi +
D_{\mu}\,dx^{\mu})^2 - e^{\psi+2\phi}\,(d\eta + C_{\mu}\,dx^{\mu}
+ \kappa\,d\chi )^2\,.
\ee
Finally, we compute
\ba
F_{\mu\nu}(C) & = & -\frac{1}{2}\sqrt{|g_{5}|}\,
e^{-\alpha\hat{\phi}}
\epsilon_{\mu\nu\lambda\rho5}\,\hat{H}^{\lambda\rho5} \nonumber
\\ & = &  \frac{1}{2}\sqrt{|g_{4}|}\,
\epsilon_{\mu\nu\lambda\rho}\,(e^{-2(\psi+\phi)}F^{\lambda\rho}(E)
- e^{-4\phi}D_{\tau} H^{\tau \lambda\rho})
\\ & = & F_{\mu\nu}(B) + F_{\mu\nu}(\kappa D)\,, \nonumber
\ea
with the definition (solving the second equation (\ref{DEpsi}))
\be
F_{\mu\nu}(B) \equiv e^{-2(\psi+\phi)}\tilde{F}_{\mu\nu}(E) -
\kappa F_{\mu\nu}(D)\,.
\ee
Accordingly we can rewrite the double Kaluza-Klein ansatz
(\ref{red64}) as
\be\lb{red64a}
ds_6^2 = e^{-\psi}\,ds_4^2 - e^{\psi-2\phi}\,\theta^2 -
e^{\psi+2\phi}\,(\zeta + \kappa\theta)^2\,,
\ee
with
\be
\theta \equiv d\chi + D_{\mu}\,dx^{\mu}\,, \qquad \zeta \equiv
d\eta + B_{\mu}\,dx^{\mu}\,.
\ee

Taking into account (\ref{an}), it follows that the ansatz for
reducing 6-dimensional vacuum gravity to EMDA may be written $$
ds_6^2 = ds_4^2 - e^{-2\phi}\,\theta^2 - e^{2\phi}\,(\zeta +
\kappa\theta)^2\,, $$
\be\lb{red64b}
\theta \equiv d\chi + \sqrt{2}A_{\mu}\,dx^{\mu}\,, \qquad \zeta
\equiv d\eta + B_{\mu}\,dx^{\mu}\,,
\ee $$ F_{\mu\nu}(B) \equiv
\sqrt{2}(e^{-2\phi}\tilde{F}_{\mu\nu}(A) - \kappa
F_{\mu\nu}(A))\,. $$

\setcounter{equation}{0}
\section{$D=6$ vacuum counterpart of BREMDA}

Using the machinery of the preceding section, we can show that BREMDA is
dual to the six-dimensional vacuum solution whose standard KK reduction gives
the usual $D=4$ dyonic BR solution with equal electric and magnetic strengths.
From (\ref{ad}) and (\ref{F})  we obtain (note that the coordinate
transformation $t,x,y,\varphi \to t,x, \theta,\varphi$ reverses orientation,
so that accordingly we must change the sign of the axion)   
\be
F_{14}(B) = 1-y^2\,, \quad F_{23}(B) = -2y\,, \quad F_{24}(B) =
-2xy\,,
\ee
leading (in a suitable gauge) to the 1-form
\be
B = -y^2\,d\varphi + x(1-y^2)\,dt\,.
\ee
It follows that the 6-dimensional line element corresponding to
(\ref{bremda}) is (with $b = -c^2$)
\ba\lb{br6a}
ds_6^2 & = & (x^2-c^2)\,dt^2  - \frac{dx^2}{x^2-c^2} -
\frac{dy^2}{1-y^2} - (1-y^2)(d\varphi + x\,dt)^2 \nonumber \\ & &
- (d\chi - xy\,dt - y\,d\varphi)^2  - (d\eta + x\,dt - y\,d\chi)^2
\,.
\ea
This may be rearranged to the more compact form
\ba\lb{br6}
ds_6^2 & = & (x^2-c^2)\,dt^2 - \frac{dx^2}{x^2-c^2} -
\frac{dy^2}{1-y^2} - (1-y^2)\,d\chi^2 \nonumber \\ & & - (d\varphi
+ x\,dt - y\,d\chi)^2 - (d\eta + x\,dt - y\,d\chi)^2 \,,
\ea
which is explicitly symmetric between the two Killing vectors
$\partial_{\varphi}$ and $\partial_{\eta}$, and enjoys the higher
symmetry group $SL(2,R) \times SO(3) \times U(1) \times U(1)$.

Eq.\ (\ref{br6}) represents the Bertotti-Robinson solution of
6-dimensional vacuum gravity. A simpler form is achieved by making
a $\pi/4$ rotation in the plane of the two Killing vectors
$(\partial_{\varphi}, \partial_{\eta})$ and relabelling the third
spacelike Killing direction according to
\be
d\varphi = -\frac{1}{\sqrt{2}}\,(d\bar{\eta} + d\bar{\chi})\,,
\quad d\eta = \frac{1}{\sqrt{2}}\,(d\bar{\eta} - d\bar{\chi})\,,
\quad d\chi = d\bar{\varphi}\,.
\ee
This leads to
\ba\lb{br51}
ds_6^2 & = & (x^2-c^2)\,dt^2  - \frac{dx^2}{x^2-c^2} -
\frac{dy^2}{1-y^2} - (1-y^2)\,d\bar{\varphi}^2 \nonumber
\\ & & - (d\bar{\chi} - \sqrt{2}x\,dt + \sqrt{2}y\,d\bar{\varphi})^2 -
d\bar{\eta}^2\,,
\ea
which is the trivial 6-dimensional embedding of the 5-dimensional
Bertotti-Robinson metric 
\ba
ds_5^2 & = & (x^2-c^2)\,dt^2  - \frac{dx^2}{x^2-c^2} -
\frac{dy^2}{1-y^2} - (1-y^2)\,d{\varphi}^2 \nonumber
\\ & & - (d{\chi} - \sqrt{2}x\,dt + \sqrt{2}y\,d{\varphi})^2 \,.
\ea
Remarkably, this is exactly the solution whose four-dimensional
Kaluza-Klein reduction is
the Einstein-Maxwell Bertotti-Robinson solution (\ref{br}), for more details
see Appendix A.

The metric (\ref{br6a}) may also be dimensionally reduced
relatively to the Killing vectors $\partial_{\eta}$ and
$\partial_t$ (instead of $\partial_{\chi}$). Choosing $c^2 = 1$,
rearranging (\ref{br6a}) as
\ba
ds_6^2 & = & (x^2-1)\,(d\varphi - y\,d\chi)^2  -
\frac{dx^2}{x^2-1} - \frac{dy^2}{1-y^2} - (1-y^2)\,d\chi^2
\nonumber
\\ & & - (dt + x\,d\varphi - xy\,d\chi)^2 - (d\eta + x\,dt - y\,d\chi)^2\,,
\ea
and relabelling the Killing directions according to $\varphi \to t
\to \chi \to -\varphi$, one obtains the equivalent 6-dimensional
metric
\ba
ds_6^2 & = & (x^2 - 1)\,(dt + y\,d\varphi)^2 - \frac{dx^2}{x^2 - 1}
- \frac{dy^2}{1 - y^2} - (1 - y^2)\,d\varphi^2 \nonumber
\\ & & -(d\chi + x\,dt + xy\,d\varphi)^2 - (d\eta + x\,d\chi + y\,d\varphi)^2\,.
\ea
Following (\ref{red64b}), this may be reduced to the solution of EMDA:
\ba\lb{brnut}
ds_4^2 & = & (x^2 - 1)\,(dt + y\,d\varphi)^2 - \frac{dx^2}{x^2 - 1}
- \frac{dy^2}{1 - y^2} - (1 - y^2)\,d\varphi^2\,, \nonumber \\
A & = & \frac{x}{\sqrt{2}}\,(dt + y\,d\varphi)\,, \qquad \phi = 0\,,
\qquad \kappa = x\,,
\ea
with $B = - x^2\,dt - (x^2-1)\,yd\varphi$ ($\kappa$ may for instance be
obtained by solving the Maxwell equations $dF(B) = 0$). This
Bertotti-Robinson-NUT solution may be obtained from the BREMDA solution
(\ref{bremda}) by the correspondence (which is an isometry of the
5-dimensional Bertotti-Robinson metric (\ref{brkk}))
\be
t \leftrightarrow i\varphi\,, \quad x \leftrightarrow -y\,, \quad ds^2
\to -ds^2\,, \quad A \to -iA\,.
\ee
The dimensional reduction of (\ref{br6}) to (\ref{brnut}) breaks the full
symmetry group of the 6-dimensional Bertotti-Robinson solution to $SO(3)
\times U(1)$.

Remarkably, this BR-NUT solution may also be obtained as a near-horizon limit
of near-extremal static black hole solutions of EMDA with NUT charge. Such
near-extremal black holes are defined by the condition that
\be
(|{\cal M}| - |{\cal D}|)^2 \equiv |{\cal M}|^2 +  |{\cal D}|^2 
- |{\cal Q}|^2 \equiv \lambda^2 c^2
\ee
is small. Putting $r - M - r_-/2 \equiv \lambda x$, we obtain for the metric
functions in (\ref{GK})
\be
\Delta =\lambda^2(x^2-c^2)\,, \qquad \Sigma = 2\lambda|{\cal M}|
\left(\frac{M}{|{\cal M}|}\,x + c\right) + \Delta \,.
\ee
So the limit $\lambda \to 0$ will yield a Bertotti-Robinson-like metric only
for $M = 0$. In this case, rescaling times by $t \to (r_0^2/\lambda)t$
as in (\ref{newcoord}), with now
\be
r_0^2 = 2\lambda N\,,
\ee
we obtain the limiting 4-dimensional metric
\be
ds^2 = r_0^2\left[\frac{x^2-c^2}{c}\,(dt + y\,d\varphi)^2 -
\frac{c}{x^2-c^2}\,(dx^2 + (x^2-c^2)\,d\Omega^2)\right]\,,
\ee
which is identical with (\ref{brnut}) after scaling $r_0^2$ to unity and
choosing without loss of generality $c = 1$ (one can always rescale
$x \to cx$ and $ds^2 \to c\,ds^2$), as this construction goes along only
for $c \neq 0$. Likewise, choosing
\be
z_{\infty} = 2iQ{\cal Q}^*/cr_0^2\,,
\ee
we obtain from (\ref{pots}) (after reversing the signs of the pseudoscalars 
$\kappa$ and $u$ as explained above) the limiting (gauge
transformed  and rescaled) scalar potentials 
\be
\phi = 0\,, \qquad \kappa = \frac{x}{c}\,, \qquad v = \frac{r_0}{\sqrt{c}}x\,, 
\qquad u = \frac{r_0}{\sqrt{c}}\,\frac{x^2-c^2}{c}\,, 
\ee
in agreement with (\ref{brnut}). Again, we note that this limit is independent 
of the original values of $Q$ and $P$, provided $Q \neq 0$.
\setcounter{equation}{0}
\section{Eleven-dimensional supergravity}

The idea to generate higher rank antisymmetric forms by dualizing the KK
two-forms was generalized to $D=11$ supergravity as follows
\cite{ChGaSh00}.
Starting with the lagrangian 
\be\label{L11d} 
S_{(11)} = \int d^{11} x \sqrt{-\hat g} \left\{ \hat R_{(11)} 
    - \frac1{2\times 4!} \hat F_{[4]}^2 \right\} 
   -\frac1{6} \int \hat F_{[4]} \wedge \hat F_{[4]} \wedge \hat A_{[3]}, 
\ee 
we use the following three-block  ansatz for
the $D=11$ metric  
\be\label{ds11} 
d \hat s_{(11)}^2 = g_2^{\frac12} \delta_{ab} dz^{a} dz^{b} 
   + g_3^{\frac13} \delta_{ij} dy^i dy^j 
  + (g_2 g_3)^{-\frac14} g_{(6)\mu\nu} dx^\mu dx^\nu, 
\ee
where all variables depend only on the six coordinates $x^\mu$ and $a,
b=1,2;\,i, j=1,2,3;\, \mu,\nu=0,...,5$. The three-form potential $A_{[3]}$ is
reduced to its six-dimensional pull-back $B_{[3]}(x)$, the one-form
$A_{[1]}=A_\mu dx^\mu$ and the scalar scalar $\kappa(x)$, with
 \be
{\hat A}_{\mu z^1 z^2}=A_{\mu}(x),\quad {\hat A}_{y^1 y^2 y^3}=\kappa (x).
\ee

For the four-form
one has
\be \label{F4} 
\hat F_{[4]} = G_{[4]} + F^2_{[2]} \wedge Vol(2) 
   + d\kappa \wedge Vol(3), 
\ee 
where $G_{[4]} = dB_{[3]}, F_{[2]} = dA_{[1]}$. After
reduction  to six dimensions we obtain the theory governed by the action
\ba\lb{6d}
S_{(6)} &=& \int d^6x \sqrt{|g_{(6)}|} \Bigl\{ R_{(6)} 
   - \frac{e^{2\phi}}2 (\nabla\kappa)^2 - \frac12 (\nabla\phi)^2 
   - \frac3{16} (\nabla\psi)^2 \nonumber\\ 
  &-& \frac12 e^{-\phi} \left[ 
     \frac1{2!} e^{\frac34 \psi}F^2_{[2]} 
   + \frac1{4!} e^{-\frac34 \psi} G^2_{[4]} \right] \Bigr\} 
   + \mbox{ Chern-Simons terms,}
 \ea
where two new scalar fields are introduced via
\be 
\ln g_2 = \frac23 \phi -  \psi, \qquad \ln g_3 = -2 \phi. 
\ee 
It is easy to see that $\kappa,\,\phi$ form a coset
$SL(2,R)/SO(1,1)$, while $F_{[2]}$ and the six-dimensional dual ${\tilde
G}_{[2]} =*G_{[4]}$ (which is also a two-form) can be combined into the
$SL(2,R)$ doublet 
\be
{\cal F}_{[2]}=e^{\psi/2} F_{[2]}+ ie^{-\psi/2} {\tilde G}_{[2]},
\ee
transforming under $SL(2,R)$ as follows
\ba  
&&z \to \frac{az+b}{cz+d}, \quad 
  z = \kappa+ie^{-\phi}, \quad 
  ad - bc = 1, \nonumber\\
&&{\cal F}_{[2]} \to (cz+d){\cal F}_{[2]}, \quad 
\psi \to \psi + \hbox{const.} 
\ea 

The multiplet of matter fields in the $D=6$ action is the same as
that which may be obtained  from compactification of $D=8$ vacuum
gravity. Moreover, the action which follows from the $D=8$ Einstein action
with the metric ansatz 
\ba
ds_{(8)}^2 &=& g_{mn} \left( d\zeta^m + A^m_{\mu} dx^\mu \right) 
    \left( d\zeta^n + A^n_{\nu} dx^\nu \right) 
  + e^{-\frac14\psi} g_{\mu\nu} dx^\mu dx^\nu,  \nonumber\\
e^{\psi} &=& \det ||g_{mn}||,   
\ea
 ($m,n=1,2$) leads exactly to the theory (\ref{6d}) after 
the identification of variables 
\ba\lb{id}
g_{mn} &=& e^{\psi/2} 
   \left( \begin{array}{cc} 
          e^\phi & \kappa e^\phi \\ 
          \kappa e^\phi & e^{-\phi}+\kappa^2 e^\phi 
          \end{array} \right), \nonumber \\ 
dA^m_{[1]} &=& F^m_{[2]}, \nonumber\\
F^1_{[2]}+\kappa F^2_{[2]}&=&e^{-\phi-\frac{3\psi}{4}}{\tilde G}_{[2]}.
\ea
More precisely, the field equation for $B_{[3]}$ becomes a Bianchi identity
for $A^1_{[1]}$ and vice versa.  Note, that interchanging the $D=8$ Killing
vectors, i.e. relabelling $A^1_\mu\leftrightarrow A^2_\mu$ we will get
different $D=11$ field configurations.
 
Now we can construct a solution to $D=11$ supergravity which is dual 
to the eight-dimensional vacuum metric obtained from BR6(\ref{br51})
smeared in two flat extra dimensions
\ba\lb{br8}
ds_8^2 & = & (x^2-c^2)\,dt^2  - \frac{dx^2}{x^2-c^2} -
\frac{dy^2}{1-y^2} - (1-y^2)\,d\varphi^2 \nonumber
\\ & & - (d\chi - \sqrt{2}x\,dt + \sqrt{2}y\,d\varphi)^2 -
d\bar{\eta}^2-dz_1^2-dz_2^2\,.
\ea
This solution possesses several commuting Killing vectors, from which
one can choose any pair to be used in KK reduction back to six dimensions.
 
Choosing  $\zeta_1=\chi, \zeta_2=\eta$,we have
\be
A^1=\sqrt{2}(yd\varphi-xdt), \quad A^2=0.
\ee
Transforming to the $D=11$ variables we obtain 
\be
g_2=g_3=1,\quad \phi=\psi=\kappa=0,\quad 
F_{[4]}=\sqrt{2}Vol(2)\wedge(dy\wedge d\varphi-
dx\wedge dt).
\ee
For the different order of vector fields $\zeta_1=\eta, \zeta_2=\chi$ one
obtains the same $g_2, g_3$ and zero scalars $\phi,\psi,\kappa$  but 
a different four-form:
\be
F_{[4]}=* \sqrt{2}(dy\wedge d\varphi-dx\wedge dt),
\ee
where a star denotes the $D=6$ Hodge dual.
In both cases the $D=11$ metric is a trivial smearing of the $D=6$ metric:
\be
ds_{11}^2=ds_6^2-dx_6^2-\ldots - dx_{10}^2.
\ee

\setcounter{equation}{0} 
\section{Breaking of supersymmetry}
Now let us discuss the issue of supersymmetry. 
As it is well-known, the Bertotti-Robinson 
solution preserves all the supersymmetries of $D=4, \,{\cal N}=2$ 
supergravity \cite{GiHu82,To83}. The Bardeen-Horowitz solution (\ref{BHz})
is a vacuum  one, so it can be probed for ${\cal N}=1$ supersymmetry. The
result is  negative: no geometric Killing spinors exist. Our solution
(\ref{nhz}) should  be tested in the context of $D=4, {\cal N}=4$
supergravity, the relevant  equations coming from the supersymmetric variation
of the dilatino and   gravitino. The variation of the dilatino leads to a
purely algebraic  equation, which in the case of a vanishing dilaton reads 
\be\lb{susy4} 
(\gamma^\mu\p\mu \kappa +i\sqrt{2} 
\sigma^{{\bar\mu}{\bar\nu}}F^-_{{\bar\mu}{\bar\nu}})\epsilon=0, 
 \ee 
where $F^-$ is the anti-self-dual part of the Maxwell tensor. 
 
Substituting here $\kappa =\cos\theta$ and the Maxwell tensor (\ref{F}) one
obtains the equation  \be 
M(\theta)\epsilon =0,\quad 
M=\gamma^{\bar\theta}\sin\theta-(\cos\theta+i)(\sigma^{\bar\theta\bar\varphi} 
-i\sigma^{{\bar t}{\bar x}}). 
\ee  
The determinant  $|\det M|=\sin^4\theta$, 
so that there is no non-trivial  solution to Eq. (\ref{susy4}), 
i.e. the BREMDA bosonic solution breaks all the supersymmetries of $D=4, 
{\cal N}=4$ supergravity. Similarly, one can show that the Bertotti-Robinson 
endowed with NUT (\ref{brnut}) is not supersymmetric either. 
 
Now discuss the $D=11$ embedding. Our $D=11$ solution is related to the
four-dimensional BREMDA
in a non-local way, since it is obtained using  dualizations
in the intermediate dimensions. So a priori it is not clear whether
it is non-supersymmetric in the supergravity sense.

It was shown \cite{ChGaSh00} that the $D=11$ Killing spinor equation 
for the 32-component Majorana 
spinor $\epsilon_{(11)}$ ensuring the vanishing of the supersymmetry 
variation of the gravitino
\be 
\hat D_M \epsilon_{(11)} + \frac1{288}\left( \Gamma_M{}^{\bar N\bar P\bar Q\bar 
R}    - 8 \delta_M^{\bar N} \Gamma^{\bar P\bar Q\bar R} \right) 
     \hat F_{\bar N\bar P\bar Q\bar R} \epsilon_{(11)} = 0 \label{Ki} 
\ee 
for the $2+3+6$ block truncation considered above corresponds to the purely
geometric equation  for the eight-dimensional dual: 
\be\lb{conspi} 
(\p{\bar \mu}-\frac14\omega^{ab}_{\bar \mu}\sigma_{ab})\epsilon_8=0 
\ee 
We will use   the flat gamma-matrices, and $a,b,\bar\mu$ are  
the tetrad and coordinate indices respectively. Here the spin-connection 
$\omega^{ab}_{\bar \mu}$ has to be calculated for the spacetime (\ref{ds11}).
 Therefore to explore the supersymmetry in the $D=11$ supergravity sense
we have to check whether the corresponding  $D=8$
solution admits covariantly constant spinors.

The non-zero spin-connection one-forms for the metric (\ref{br8}) read 
(we use tetrad indices and numbering $0,1,2,3,4$ for
$t,\xi,\theta,\varphi,\chi)$:
\ba
&&\omega_{01}=\cos\theta d\varphi+d\chi/\sqrt{2}  ,
\quad \omega_{04}= d\xi/\sqrt{2} ,
\quad \omega_{14}= -\sinh\xi dt/\sqrt{2} ,\nonumber\\
&&\omega_{23}=\cosh \xi dt - d\chi/\sqrt{2}  ,
\quad \omega_{24}= -\sin\theta d\varphi/\sqrt{2} ,
\quad \omega_{34}= d\theta/\sqrt{2} .
\ea
For the gamma matrices in eight dimensions one can use suitably defined
tensor products of Pauli matrices. Since the spin-connection lies
entirely in the $D=6$ sector, one can suppress spinor indices relating
to the transition from six to eight dimensions and use $8\times 8$
gamma matrices and $D=6$ spinors. A convenient choice is
\ba
&&\Gamma^0=i\sigma_1 \otimes 1\otimes 1,\quad
\Gamma^1=\sigma_2 \otimes 1\otimes 1,\nonumber\\
&&\Gamma^2=\sigma_3 \otimes \sigma_1\otimes 1,\quad
\Gamma^3=\sigma_3 \otimes \sigma_2\otimes 1,\nonumber\\
&&\Gamma^4=\sigma_3 \otimes \sigma_3\otimes\sigma_1,\quad
\Gamma^5=\sigma_3 \otimes\sigma_3\otimes\sigma_2.
\ea
The corresponding Lorentz generators are:
\ba
&&\sigma^{01}=-\sigma_3\otimes 1\otimes 1,\quad
\sigma^{04}=\sigma_2\otimes \sigma_3\otimes \sigma_1,\nonumber\\
&&\sigma^{14}=i\sigma_1\otimes \sigma_2\otimes\sigma_1,\quad
\sigma^{23}=i 1\otimes\sigma_3\otimes 1,\nonumber\\
&&\sigma^{24}=-i 1\otimes\sigma_2\otimes\sigma_1,\quad
\sigma^{34}=i 1\otimes\sigma_1\otimes \sigma_1.
\ea
A direct substitution in the Eq.(\ref{conspi}) gives a system of matrix
equations which should satisfy the integrability conditions
\be
R^{ab\mu\nu }\sigma_{ab}\epsilon_6=0
\ee
where the mixed coordinate-tetrad components of the $D=6$ Riemann tensor
are introduced. Writing them as curvature two-forms $\Omega^{ab}=
R^{ab}_{\mu\nu }dx^\mu\wedge dx^\nu$, one finds the following non-zero
quantities
\ba
&&\Omega^{01}=\sin\theta d\theta\wedge d\varphi-
\frac12\sinh\xi dt\wedge d\xi,\quad 
 \Omega^{02}=\frac12\sin\theta d\xi\wedge d\varphi,\nonumber\\
&&\Omega^{03}=\frac12 d\theta\wedge d\xi,\quad
 \Omega^{04}=\frac{1}{\sqrt{2}}\cos\theta\sinh\xi dt\wedge d\varphi
+\frac12\sinh\xi dt\wedge d\chi, \nonumber\\
&&\Omega^{12}=\frac12\sin\theta \sinh\xi dt\wedge d\varphi,  \quad
\Omega^{13}=\frac12 \sinh\xi d\theta\wedge dt, \nonumber\\  
&&\Omega^{14}=\frac{1}{\sqrt{2}}\cosh\xi dt\wedge d\xi
+\frac{1}{\sqrt{2}} \cos\theta d \xi\wedge d\varphi
+\frac12 d\xi\wedge d\chi,\nonumber\\ 
&&\Omega^{23}=-\sinh\xi dt\wedge d\xi+
\frac12\sin\theta d\theta\wedge d\varphi,\nonumber\\
&&\Omega^{24}=-\frac{1}{\sqrt{2}}\cosh\xi dt\wedge d\theta-
 \frac{1}{\sqrt{2}}\cos\theta d \theta\wedge d\varphi
 -\frac12 d\theta\wedge d\chi,\nonumber\\
&&\Omega^{34}=-\frac{1}{\sqrt{2}}\sin\theta\cosh\xi dt\wedge d\varphi-
\frac12\sin\theta d\varphi\wedge d\chi. 
\ea
All ten two-forms are independent, so one obtains ten integrability
conditions:
\ba
&&\sigma^{03}\epsilon_6=\sigma^{14}\epsilon_6=\sigma^{24}\epsilon_6=
\sigma^{34}\epsilon_6=0, \label{sieps} \\
&&(\tanh\xi\sigma^{01}+\sqrt{2}\sigma^{14})\epsilon_6=0,\quad
(\tanh\xi\sigma^{13}+\sqrt{2}\sigma^{24})\epsilon_6=0, \nonumber\\
&&(\cot\theta\sigma^{04}-\coth\xi\sigma^{04}+\sqrt{2}\sigma^{12})
\epsilon_6=0,\nonumber\\
&&(\sigma^{04}+{2}\sigma^{23})\epsilon_6=0,\quad
(\tan\theta\sigma^{02}-\sqrt{2}\sigma^{14})\epsilon_6=0, \nonumber\\
&&( 2\sigma^{01}-\sigma^{23}+\sqrt{2}\cot\theta\sigma^{24})
\epsilon_6=0.
\ea
These  are clearly inconsistent (inconsistent are  already
 conditions (\ref{sieps}), since sigma-matrices do not have kernels).
Therefore the BREMDA geometry is not supersymmetric in the sense of
$D=11$ supergravity either.

\section{Conclusion}

We have presented a new solution to dilaton-axion gravity in four dimensions
whichis a rotating version of the Bertotti-Robinson metric. It
breaks the $SO(3)$ symmetry of the latter but preserves the $SL(2,R)$
symmetry of the anti-de Sitter sector. The metric arises as the near-horizon
limit of the charged rotating axion-dilaton black hole (in the theory with one
vector field) and  is supported  by  non-trivial vector and 
axion fields. It looks simpler than the  near-horizon Kerr (or Kerr-Newman)
metric due to the absence of additional angular factors, while preserving
the same mixing of the azimuthal and time coordinates induced by rotation.
It is important to note that the AdS sector does not
factor out even asymptotically. Moreover, in contrast to  the case of
$AdS_2\times S^2$, the conformal boundary is now a singular $1+2$ space.

The new metric was shown to be related to the usual dyonic BR solution
with equal electric and magnetic charges after uplifting it to six dimensions
and then coming back along a different reduction scheme. In this procedure 
the axion emerges via dualization of one of the Kaluza-Klein two-forms.
Using a similar reasoning, we were able to find two solutions of $D=11$
supergravity with non-trivial four-form fields whose dimensional reduction
(including dualizations at intermediate steps) gives our solution. 
 
This solution is not supersymmetric, whether in the sense of the
original $D=4, {\cal N} =4$ supergravity, or in the sense of higher
dimensional embeddings. But it still looks promising from the point of view of
holography. Indeed, it preserves some features of $AdS_2\times S^2$ found
also for the Kerr throat \cite{BaHo99}, and it is not plagued by superradiance
as the latter. Preliminary considerations show that, in spite of the singular
nature of the boundary, the asymptotic symmetry contains the Virasoro
algebra \cite{ClGa01}. Also it is likely to provide a new version of 
conformal mechanics of the type studied recently \cite{DFF,Cl98}. We will
discuss these issues in a separate publication.  
 
\section*{Acknowledgements} 
We would like to thank K. Bronnikov, J. Fabris, L. Palacios, and S. Solodukhin
for discussions. DG is grateful 
to LAPTH Annecy for hospitality and  to CNRS for support which made
possible this collaboration. 
His work was also supported in part by the RFBR
grant 00-02-16306.   
\section*{Appendix A}
\def\theequation{A.\arabic{equation}}
\setcounter{equation}{0}
In this Appendix we discuss the near-horizon limit to static black
hole solutions of 5-dimensional sourceless Kaluza-Klein theory, i.e.
5-dimensional vacuum Einstein gravity
\be
S = - \int d^5x \sqrt{|g_5|} R_5 \,,
\ee
together with the assumption of a spacelike Killing vector $\partial/\partial
x^5$. The 5-dimensional metric may be reduced to 4 dimensions
by the Kaluza-Klein dimensional reduction
\be\lb{red4}
ds_5^2 = e^{-2\sigma/\sqrt{3}}\,ds_4^2 -
e^{4\sigma/\sqrt{3}}\,(dx^5 + 2A_{\mu}dx^{\mu})^2\,,
\ee
with the reduced action
\be
S = \int d^4x\sqrt{|g_4|}\left\{-R_4 +
2\partial_\mu\sigma\partial^\mu\sigma
-e^{2\sigma/\sqrt{3}}F_{\mu\nu}F^{\mu\nu}\right\} \,.
\ee

The general static, NUT-less black hole solution of Kaluza-Klein
theory was derived by Gibbons and Wiltshire \cite{GiWi86}, and
generalized to rotating black hole solutions by Rasheed
\cite{Ra95}. We will consider only static black holes, which
depend on 3 parameters $M$ (mass), $\Sigma$ (scalar charge), $Q$
and $P$ (``electric'' and ``magnetic'' charge) constrained by
\be
\frac{Q^2}{\Sigma + M/\sqrt{3}} + \frac{P^2}{\Sigma - M/\sqrt{3}} = \frac{
2\Sigma}{3}\,.
\ee
The corresponding 5-dimensional metrics, as well as their 4-dimensional
reductions, have two regular horizons provided $M^2 +\Sigma^2 - P^2 - Q^2 \ge
0$. The condition of extremality is therefore $M^2 +\Sigma^2 = P^2 + Q^2$.
However for more generality we shall consider near-extremal black holes, with
\be
M^2 +\Sigma^2 - P^2 - Q^2 \equiv \lambda^2c^2
\ee
small.

The black hole solutions are
\ba\lb{bhkk}
ds_5^2 & = & \frac{f^2}{B}\,dt^2 - A\,(\frac{dr^2}{f^2} + d\theta^2 + \sin^2
\theta\,d\varphi^2) \nonumber \\
& & - \frac{B}{A}\,(dx^5 + \frac{2Q}{B}(r-M+M_-)\,dt + 2P\cos\theta\,
d\varphi)^2\,,
\ea
where the metric functions $f$, $A$, $B$ are given by
\ba
f^2 & = & (r - M)^2 - \lambda^2c^2 \,, \nonumber \\
A & = & f^2 + 2M_-(r-M) + \frac{M_-}{M}(M_+M_- + \lambda^2c^2)\,, \\
B & = & f^2 + 2M_+(r-M) + \frac{M_+}{M}(M_+M_- + \lambda^2c^2)\,, \nonumber
\ea
and
\be
M_{\pm} \equiv M\pm \frac{\Sigma}{\sqrt{3}}\,,
\quad Q^2 = \frac{M_+(M_+^2-\lambda^2c^2)}{2M}\,,
\quad P^2 = \frac{M_-(M_-^2-\lambda^2c^2)}{2M}\,.
\ee

Putting $r-M \equiv \lambda x$,  we shall take the near-extremal,
near-horizon limit $\lambda \to 0$ such that the two horizons $r = r_{\pm}
\equiv M \pm \lambda c$ approach each other while the radial coordinate
approaches the event horizon $r_+$. Four-dimensional sections $\varphi =$
const. of the 5-dimensional metric (\ref{bhkk}) being similar in form to
the rotating metric (\ref{extrds}), with the electric potential $A_4$ playing
the part of the angular velocity, to obtain a finite limit we again must
first transform to a frame ``nearly co-rotating'' with the horizon, through
a gauge transformation
\be
d\ol{x}^5 = dx^5 + \frac{2QM_-}{B(0)}\,dt\,,
\ee
with $B(0) \equiv B(r-M=0)$, leading to the ``electric'' field in
the new gauge
\be
\overline{A}_4 = -\frac{QM_-}{B(0)B}(r-M)(\frac{M_+M_-}{M} + r-M +
O(\lambda^2c^2))\,,
\ee
and rescale both time and the fifth coordinate, through the transformations
\be
r-M \equiv \lambda x\,, \quad \cos\theta \equiv y\,, \quad t
\equiv \frac{\sqrt{A_0B_0}}{\lambda}\ol{t}\,, \quad \ol{x}^5
\equiv \sqrt{2}P\chi\,,
\ee
where $(A_0,B_0) = \lim_{(\lambda\to 0)}(A,B)$. Taking the limit
$\lambda \to 0$, using the identities
\be
\frac{2B_0P^2}{A_0^2} = \frac{Q^2}{P^2}\frac{A_0}{B_0}\frac{M_-^2}{M_+^2}
= 1\,,
\ee
and relabelling the time coordinate $\ol{t} \to t$, we finally obtain the
5-dimensional near-horizon metric
\be\lb{brkk}
A_0^{-2}ds_5^2 = (x^2-c^2)\,dt^2  - \frac{dx^2}{x^2-c^2} -
\frac{dy^2}{1-y^2} - (1-y^2)\,d\varphi^2 - (d\chi - \sqrt{2}x\,dt
+ \sqrt{2}y\,d\varphi)^2 \,.
\ee
Again, as in the case of EMDA, all the static Kaluza-Klein black
holes have the same near-horizon limit, independently of the
values of the electric and magnetic charges $Q \neq 0$ and $P \neq
0$.

In (\ref{brkk}) we recognize the Kaluza-Klein version of the
dyonic Bertotti-Robinson solution with equal electric and magnetic
charges. The vanishing of the Kaluza-Klein scalar field $\sigma$
is due to the fact that the electric and magnetic fields
\be
F_{14} = - F_{23} = - \frac{1}{\sqrt{2}}
\ee
being equal in magnitude, the source term in the scalar field
equation
\be
\nabla^2\sigma =
-\frac{1}{2\sqrt{3}}e^{2\sqrt{3}\sigma}F^{\mu\nu}F_{\mu\nu}
\ee
vanishes. Accordingly the isometry group is the direct product of
that of the Bertotti-Robinson spacetime with the Klein circle,
{\sl i.e.} $SL(2,R)\times SO(3) \times U(1)$.

\end{document}